# Complex network analysis of pore structures in monodisperse granular materials with varied grain shapes

Author 1
Jie Qi, MSc (CE)
Department of Infrastructure Engineering, The University of Melbourne, Parkville, Australia
ORC ID: 0000-0002-9189-7620

Author 2
Wenbin Fei, PhD
College of Civil Engineering, Hunan University, Changsha, Hunan 410082, PRC
ORCID: 0000-0002-9275-8403
Email: wenbinfei@outlook.com

Author 3
Guillermo A. Narsilio, PhD, MSc (Math), MSc (CE), CEng
Department of Infrastructure Engineering, The University of Melbourne, Parkville, Australia
ORC ID: 0000-0003-1219-5661

*Abstract:*
Understanding how pore structure influences flow and transport behaviour in granular materials is essential for addressing a wide range of geotechnical, hydraulic, and environmental challenges. These processes are largely shaped by the microscopic arrangement of particles and interconnections between pores within the material. However, detailed insights considering granular assemblies with diversified grain shapes remain scarce. This study introduces a comprehensive framework incorporating Discrete Element Method (DEM) simulations, image processing, pore-network modelling, and complex network theory to investigate the links between particle morphology and their hydraulic behaviours. Mono-disperse assemblies of natural sand particles with varied shapes are constructed in DEM, and pore networks are extracted through image processing and pore-network modelling. Complex network analysis is then applied to calculate structural metrics that reveal intrinsic relationships between pore microstructures and hydraulic properties. Our results demonstrate that particle morphology significantly impacts pore network characteristics, including pore and throat sizes, closeness centrality, pore structure anisotropy, providing valuable insights into how pore structure influence transport properties.

*Keywords*
complex network theory; pore network; granular material; hydraulic conductivity; particle shape

## List of Notations

| Symbol | Description |
|---|---|
| $A$ | Cross-sectional area of sample perpendicular to the flow direction. |
| $A_{ti}$ | Area of the i'th throat. |
| $C$ | Conductance of the pore-throat-pore system. |
| $C_B$ | Betweenness centrality of networks. |
| $C_C$ | Closeness centrality of networks. |
| $C_{ij}$ | The hydraulic conductance the conduit between pores $i$ and $j$. |
| $C_{pk}$ | Conductance of the pore conduit k. |
| $C_{teqv}$ | Equivalent conductance of pore throats in the network. |
| $C_{ti}$ | Conductance of the conduit $i$. |
| $Cw_C$ | Weighted closeness centrality of networks. |
| $d_e$ | Equivalent diameter of particles. |
| $dh$ | The incremental distance for rendering. |
| $D_P$ | Equivalent diameter of pores. |
| $D_T$ | Equivalent diameter of throats. |
| $K$ | The permeability. |
| $L$ | Length of the sample in the direction of flow. |
| $L_A$ | Dimension of the granular assembly along rendering direction. |
| $L_e$ | The effective flow path length through the pore network |
| $L_d$ | The direct length between the inlet and outlet. |
| $l_{ij}$ | Length of the throat connecting pores $i$ and $j$. |
| $L_{pk}$ | Length of the conduit k. |
| $L_{ti}$ | Length of the conduit i. |
| $L_{teqv}$ | Equivalent length of the pore throats in the network. |
| $L_{p1}$ | Length of pore p1. |
| $L_{p2}$ | Length of pore p2. |
| $L_T$ | Length of throats. |
| $N_P$ | Number of pixels of rendered images. |
| $N_t$ | Total number of throats connecting the two pores. |
| $P_i, P_j$ | The pressures at pores $i$ and $j$. |
| $Q$ | Volumetric flow rate. |
| $Q_{ij}$ | The flow rate between pores $i$ and $j$. |
| $R$ | Roundness of particles. |
| $r_{pk}$ | Radius of the pore conduit k. |
| $r_{ti}$ | Radius of the throat conduit i. |
| $S$ | Sphericity of particles. |
| $T$ | The tortuosity. |
| $\Delta P$ | Pressure difference applied across the sample. |
| $\Delta p$ | Local pressure difference between pores driving the flow. |
| $\eta$ | Dynamic viscosity of the fluid. |

# 1  Introduction

The flow and transport phenomena in granular materials are widely encountered in various scientific and engineering systems. Understanding these processes has significant implications in fields such as geothermal exploitation (Bidarmaghz et al., 2021), $CO_2$ geo-storage (Fei et al., 2015), offshore geotechnics (Randolph et al., 2011), and radioactive waste disposal (Alonso et al., 2024). These flow and transport phenomena are fundamentally influenced by the microstructures within the granular materials (Van der Linden, 2019). However, the understanding of the relationship between these phenomena and the microstructural characteristics, particularly particle shape (Fei & Narsilio, 2020) and pore structures (Van der Linden et al., 2018), is still limited.

Advancements in imaging techniques such as X-ray computed micro-tomography (μCT), Nuclear Magnetic Resonance (NMR), and Scanning Electron Microscopy (SEM) have enabled detailed characterisation of pore space information in granular assemblies. Among them, μCT is a no-invasive imaging technique that projects X-rays through a sample from different angles and combining the detected absorption levels to form detailed 3D images of the internal structure of the granular materials (Blunt et al., 2013; Mukunoki et al., 2016; Lv et al., 2020). To achieve a high resolution, μCT scanning is usually applied to small samples. NMR imaging uses a magnetic field and radio waves to align and then knock atoms out of alignment. As they realign, they release energy, which is detected to create detailed internal images. NMR enables the imaging of geo-materials at a larger scale (Kantzas et al., 2005; Minagawa et al., 2008). SEM uses a focused beam of electrons to scan a sample's surface, producing high-resolution 2D images of topography and composition (Prakongkep et al., 2010). However, it is a destructive technique, as the electron beam can damage or alter the sample.

Pore-scale numerical simulation of flow and transport phenomena in granular materials has achieved significant success through various numerical methods. Grid-based Computational Fluid Dynamics (CFD), including Finite-Difference (FDM), Finite-Volume (FVM), and Finite-Element (FVM), solve the Navier-Stokes (N-S) equations in discretised space, are widely adopted in pore-scale fluid flow modelling (Icardi et al., 2014; Zheng et al., 2021; Zhao & O'Sullivan, 2022). However, these methods rely heavily on the quality of meshes, which can affect the accuracy and stability of simulations. Instead of solving the N-S equations directly, the Lattice Boltzmann Method (LBM) tracks the probability of fluid particles moving in specific directions at the mesoscale (Mohamad, 2019). This method is particularly effective in handling complex geometries and multiphase flows in granular materials, making it a valuable tool in various pore-scale simulations (Sun et al., 2013; Ding & Xu, 2018; Li et al., 2019). Smoothed Particle Hydrodynamics (SPH) is a mesh-free method that represents fluids using a set of particles and solves the Navier-Stokes equations in a Lagrangian framework. Such nature makes SPH highly adaptable to the complex and irregular geometries typical of pore spaces in granular materials (Zhu et al., 1999; Tartakovsky et al., 2008).

Pore network modelling is a computationally effective approach for understanding pore structures of granular materials and the associated transport behaviours. This approach simplifies the complicated pore structures into graph-like structures composing nodes and edges. Different methods have been employed to extract the pore network model from granular materials. The Medial Axis method constructs a lower-dimensional skeleton of the pore space by thinning it to a central framework, which is identified using a distance transform and ridge extraction. This skeleton is then represented as a graph of nodes and edges, simplifying the complex geometry (Liang et al., 2019; Lindquist et al., 1996). The maximal ball approach inscribes spheres within pore walls, centred at the points farthest from the particles, to extract pores and throats. This method commonly assumes specific shapes for pore space features due to the inherent complexity of the pore structure(Gerke et al., 2020). Delaunay Tessellation segments the pore space using tetrahedrons formed with vertices at the centres of neighbouring particles. It is highly effective for densely packed, spherical particles (Van der Linden et al., 2018; Sufian et al., 2019) but encounters difficulties handling angular or irregular structures. The Watershed algorithm has been applied to segment pores and throats in 3D images of porous materials by computing distance transform on binarized 3D images. This method is highly adaptable to complex and irregular structures (Rabbani et al., 2016; Taylor et al., 2015). Versatile tools based on the watershed such as OpenPNM (J. Gostick et al., 2016) and PoreSpy ( J. T. Gostick, 2017; J. Gostick et al., 2019) have been developed to extract pore networks from CT images. Roy et al., (2023) proposed a geometrical-based approach to segment pore spaces without predefined shapes, enhancing flexibility and accuracy in characterising pore structures. These methods collectively provide powerful tools for analysing and understanding the pore structures and transport behaviours in granular materials.

Despite the success achieved by imaging techniques, simulations and pore network modelling, the understanding of the intrinsic links between pore structures and transport behaviours is still limited. Complex network theory bridges the gap by quantifying the connectivity and topology of pore networks with rigorous metrics. Russell et al. (2016) explored the evolution of grain fabric and associated pore space in spherical granular media using complex networks, focusing on pore connectivity and optimized flows in the presence of shear bands. Van Der Linden et al. (2016) explored the relationship between the permeability of spherical assemblies and the complex network features of pore networks weighted by local hydraulic conductance. Zhang et al. (2022) investigated the influences of pore structure on the flow characteristics of landslide materials (modelled by spherical DEM particles) through the application of complex network theory. However, the complex-network-theory-based understanding of flow and transport behaviours in more complicated granular systems, especially those with irregular grain shapes, remains inadequately explored.

This research investigates the influence of pore structures on the hydraulic behaviours of mono-disperse granular materials with varied grain shapes using complex network analysis. Mono-sized granular assemblies were first generated using the general-shape DEM. Three-dimensional (3D) image stacks of the pore space were extracted with in-house Python codes. Pore networks were then extracted from these 3D images using OpenPNM and were weighted by local hydraulic conductance. The geometry properties of the pore networks, along with complex network features, were analysed to provide insights into the influence of particle shape. The outcomes of this research have significant potential for advancing more complex studies and carry important implications for engineering applications such as erosion prevention in earth dams, underground seepage control, and soil improvement techniques.

## 2 Methodology

The methodology employed in this research involves a comprehensive framework designed to investigate the influence of particle shape on the hydraulic properties of granular materials. Initially, irregularly shaped particle assemblies were generated using the general-shaped DEM. Blender was then used to slice these three-dimensional assemblies, creating image stacks that were subsequently processed with ImageJ to produce binary representations of the pore spaces. These binary images were segmented using the SNOW algorithm based on PoreSpy and converted into pore network models using OpenPNM. Complex network theory was applied to analyse the pore networks with local hydraulic conductance used as weights, focusing on metrics such as closeness and betweenness centrality to gain insights into the connectivity and fluid flow pathways within the granular assemblies.

### 2.1 Irregular particles

Particle shape is a crucial factor influencing the flow and conduction behaviours in granular materials. In this research, detailed particle geometry of natural sand particles was obtained through CT scanning (Fei et al., 2019) in STL format. The STL meshes were then simplified to reduce the number of facets thus saving computational resources while keeping the overall morphology. The simplification was achieved based on the Quadric Edge Collapse Decimation algorithm (Garland & Heckbert, 1997), implemented in MeshLab (Cignoni et al., 2008). This method effectively reduces the number of faces by collapsing edges according to a cost function, which minimizes the deviation from the original mesh geometry. Following simplification, all particles were uniformly scaled to an equivalent diameter of $d_e = 3mm$.

Various shape parameters were applied to characterize particle shape(Fei et al., 2021). Sphericity (S) measures particle shape at the diameter scale, indicating how closely the particle approximates a spherical shape. Meanwhile, roundness (R) quantifies corner sharpness, assessing the extent of particle corner rounded:

$$S = \frac{36\pi V_P^2}{SA^3} \quad (1)$$

$$R = \frac{\sum r_i/N_{cor}}{r_{in}} \quad (2)$$

where $V_P$ represents particle volume; SA is the surface area; $r_i$ is the radius of the *i'th* corner; $N_{cor}$ is the total number of corners, $r_{in}$ is the maximum inscribed radius.

Using the open-source code MechSys (Galindo-Torres, 2013), general-shape DEM models based on sphero-polyhera were developed to create irregular particles (Table 1). The particles were named (G1 to G5) based on the sequence of sphericity.

*Table 1 Sand particle geometry based on CT scanning*

| ID | G0 | G1 | G2 | G3 | G4 | G5 |
|---|---|---|---|---|---|---|
| Sphericity | 1.00 | 0.90 | 0.86 | 0.79 | 0.72 | 0.59 |
| Shape | 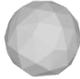 | 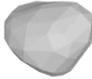 | 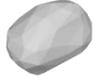 | 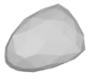 | 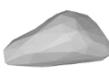 | 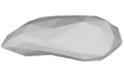 |

## 2.2 General-shape DEM assemblies

The irregular particles from natural sand were used to construct DEM assemblies within a rigid specimen with a dimension of X×Y×Z = 60×36×36 mm. Each assembly contains only one particle shape with an equivalent diameter of $d_e$=3 mm. The sphero-radius ($S_r$) for edges and corners was set to 0.05 mm (1/60 $d_e$) to achieve a reasonable representation of the particles (Galindo-Torres, 2013). Particles were arranged in a 5x5 grid sequentially at the top with random orientations and settled to the bottom under gravity (Figure 1 (a)). New layers were generated after the previous layer stabilized until the desired particle number was reached. Compaction was then conducted after the generated particle stabilised by applying a plate at 0.01 mm/s from the sample top until achieving a target porosity of 0.4 (Figure 1(b)). After stabilization, the particles were then fixed to form a rigid structure. The DEM simulation parameters are provided in Table 2.

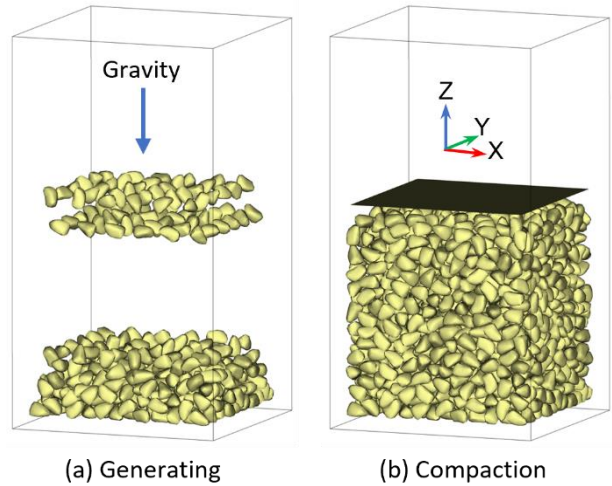

*Figure 1 General-shape granular assemblies (a) DEM particle generating; (b) assembly compaction*

*Table 2 Summary of simulation parameters*

| Parameters | Values |
| --- | --- |
| Equivalent diameter | 3mm |
| Particle density | 2500kg/m$^3$ |
| Restitution coefficient, | 0.55 |
| Friction coefficient | 0.31 |
| Sphero radius | 0.05mm |
| DEM simulation time step | $2\times10^{-6}$s |
| DEM sample dimension | 36mm |
| Number of DEM particles | 1945 |

## 2.3 Pore space 3D images

Extracting the detailed pore structure from the granular assembly is a fundamental step to characterise the pore networks. With the geometry of generated granular assemblies obtained, the geometries of the pore space with dimensions of 30×30×30mm were generated through Boolean operations and stored as STL meshes, as depicted in Figure 2. The variability in particle geometry leads to unique pore distributions within each assembly. The STL meshes of these assemblies allow for precise characterisation of the pore networks, facilitating advanced analysis of the hydraulic properties and internal erosion patterns in granular materials.

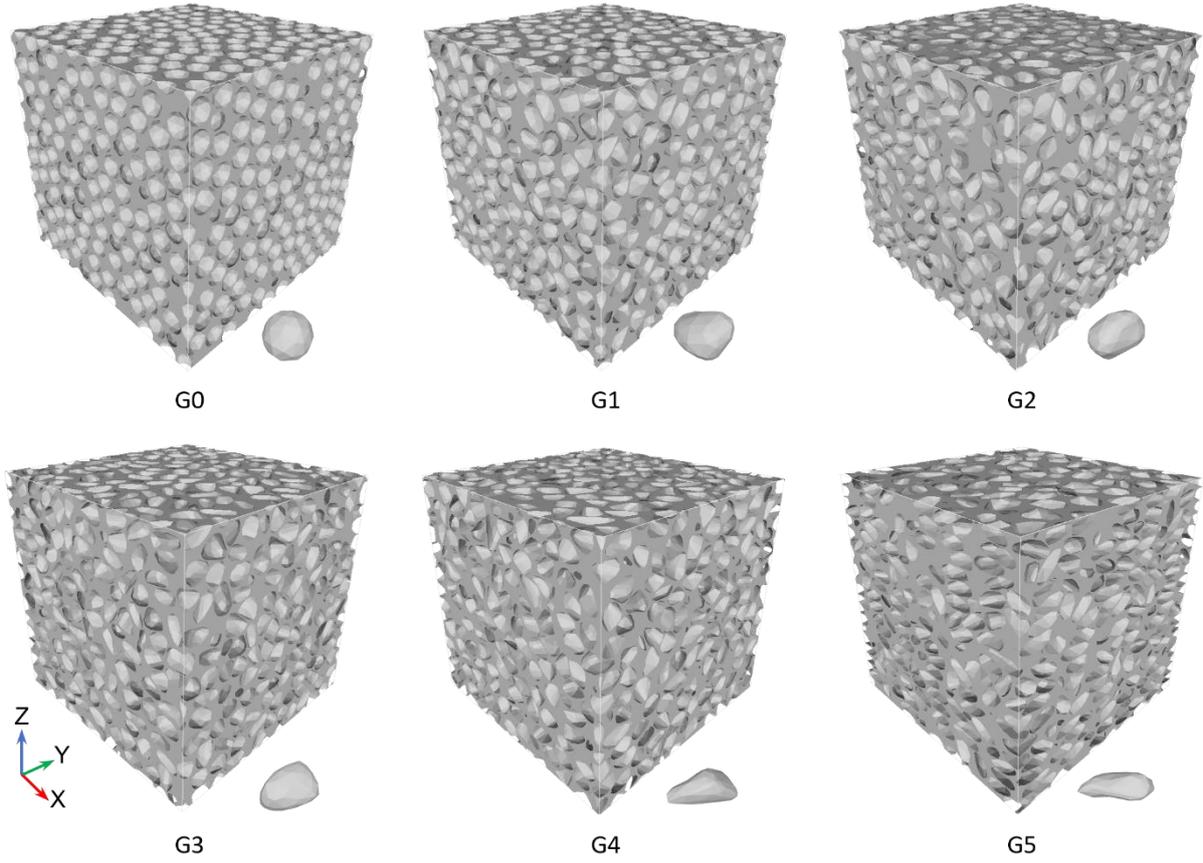

*Figure 2 Pore geometry of general-shape granular assemblies*

The 2D greyscale images of the pore space were obtained using Blender (Blain, 2019). A Python plugin script embedded in Blender was developed to create the 2D greyscale image of the slices. The process begins by importing a granular assembly model in STL format into Blender with correct positioning. A plane was then created and scaled to intersect the granular assembly model horizontally, generating cross-sectional geometry of the pore space through Boolean operations. The cross-sectional geometry was then assigned a distinct material colour to enhance visibility. Subsequently, a "camera" was added in the orthographic direction of the planes. The distance between the camera and the plane was fixed to ensure the proper size and position of the cross-sections in the images. High-quality rendering was then conducted to create grayscale images of the intersection Figure 3 (b). The plane and camera and moved along the Z axis sequentially with an incremental distance of *dh* to scan through the assembly and create images at different cross-sections. The distance between two layers is determined by the dimension of the sample and the desired resolution of images:

$$dh = \frac{L_A}{N_P - 1} \tag{3}$$

where *dh* the incremental distance for rendering, $L_A$ dimension of the granular assembly along the z direction, $N_P$ number of pixels of created images.

ImageJ (Abràmoff et al., 2004) was employed to convert the greyscale 2D images to binary images and create 3D image stacks in the sequence of the slicing procedure, as shown in Figure 3 (c). The binary conversion process in ImageJ involves thresholding techniques, where the pores were represented by white pixels with a value of 255 and the solid matrix by black pixels with a value of 0. After the conversion, the binary images were aligned sequentially to create a 3D image stack that represents the pore network in the entire granular assembly.

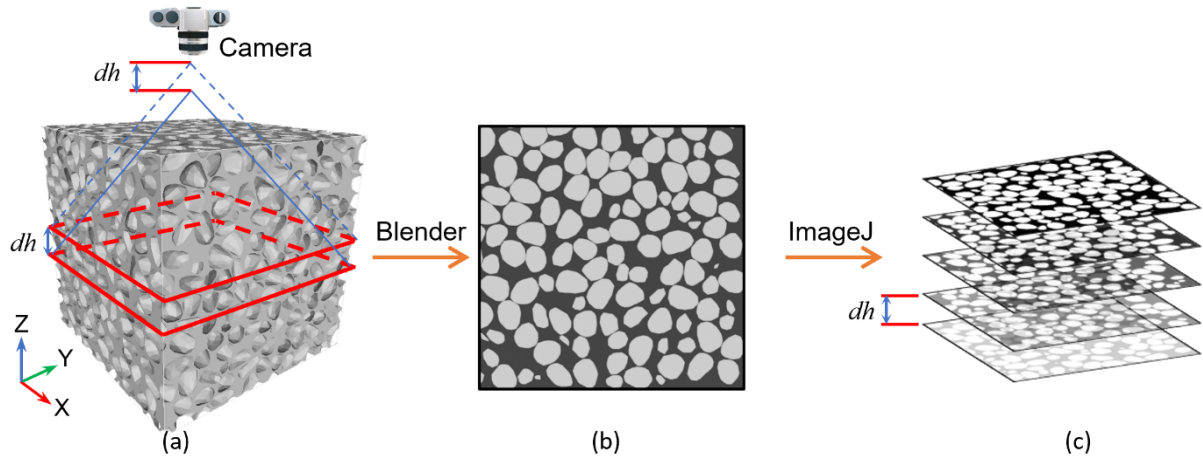

*Figure 3 Pore image extraction (a) pore geometry; (b) greyscale images of 2D pore slices; (c) 3D binary pore image stacks*

## 2.4 Pore-network construction

To construct the pore network models, binary images of granular assemblies (Figure 4 (a)) were first processed using PoreSpy (J. Gostick et al., 2019), a Python-based toolkit for the analysis of porous media. Based on the 3D image stacks, the SNOW (Sub-Network of Watersheds) algorithm (J. T. Gostick, 2017) was employed for pore space segmentation. The SNOW algorithm is an advanced method based on the watershed segmentation technique, which is particularly effective for identifying and separating connected regions in an image. This process begins with computing the distance transform of the binary image stack, which measures the distance from each pixel in the pore space to the nearest solid boundary Figure 4 (b). Local maxima of the distance transform were identified as markers, representing the centres of pores. The watershed algorithm then floods the distance map from these markers, segmenting the pore space into individual pores and defining the throats (boundaries between pores). This segmentation process effectively characterises the pore network, providing detailed information about the structure and connectivity of the pores and throats.

Based on the SNOW segmentation, geometric parameters of the segmented pores and connecting throats were extracted, as shown in Table 3. Detailed information on pores, including volume, coordinates, geometric centroid, and inscribed diameter, as well as throats, including connections, cross-sectional area, length, and inscribed diameter, was recorded for further pore network construction and analysis. The number of valid values provided in the table represents the total count of data points available for each property, derived from the G1 assembly as an example in Table 3.

*Table 3 Pore network geometric properties*

| Number | Properties | Description | Number of Valid Values (G1) |
|---|---|---|---|
| 1 | pore.area | Cross-sectional area of the pores | 5516 |
| 2 | pore.conductance | Conductance of the pores | 5516 |
| 3 | pore.coords | Coordinates of the pores | 5516 |

| 4 | pore.equivalent_diameter | Equivalent diameter of the pores | 5516 |
|---|---|---|---|
| 5 | pore.extended_diameter | Extended diameter of the pores | 5516 |
| 6 | pore.geometric_centroid | Geometric centroid of the pores | 5516 |
| 7 | pore.inscribed_diameter | Inscribed diameter of the pores | 5516 |
| 8 | pore.surface_area | Surface area of the pores | 5516 |
| 9 | pore.volume | Volume of the pores | 5516 |
| 10 | throat.conns | Connections between the throats | 16298 |
| 11 | throat.cross_sectional_area | Cross-sectional area of the throats | 16298 |
| 12 | throat.direct_length | Direct length of the throats | 16298 |
| 13 | throat.equivalent_diameter | Equivalent diameter of the throats | 16298 |
| 14 | throat.inscribed_diameter | Inscribed diameter of the throats | 16298 |
| 15 | throat.total_length | Total length of the throats | 16298 |
| 16 | throat.volume | Volume of the throats | 16298 |

The segmented data from the SNOW algorithm, including pore coordinates and throat connections, was used to create a pore network model in OpenPNM (J. Gostick et al., 2016), an open-source framework designed for pore network modelling. For pore network model construction, each pore within the network was represented by its maximum inscribed sphere, providing a geometrically simplified yet accurate representation of the pore spaces. Meanwhile, the throats, which connect adjacent pores, were modelled as cylinders as depicted in Figure 3(c). By employing these geometric representations, the resultant three-dimensional pore network model captures the spatial distribution of pores/throats and their connectivity within the assembly.

To facilitate visualization of the model, the constructed pore network models were exported in Visualization Toolkit (VTK) format(Schroeder et al., 1998) for 3D computer graphics, image processing, and visualization. The visualization as depicted in Figure 3(d) was achieved in Paraview (Ahrens et al., 2005), an open-source, multi-platform data analysis and visualization application. This comprehensive extraction and modelling approach allow for a detailed analysis of the structural and hydraulic properties of the material, providing insights into how particle shape and size distribution impact the overall functionality of the granular assemblies.

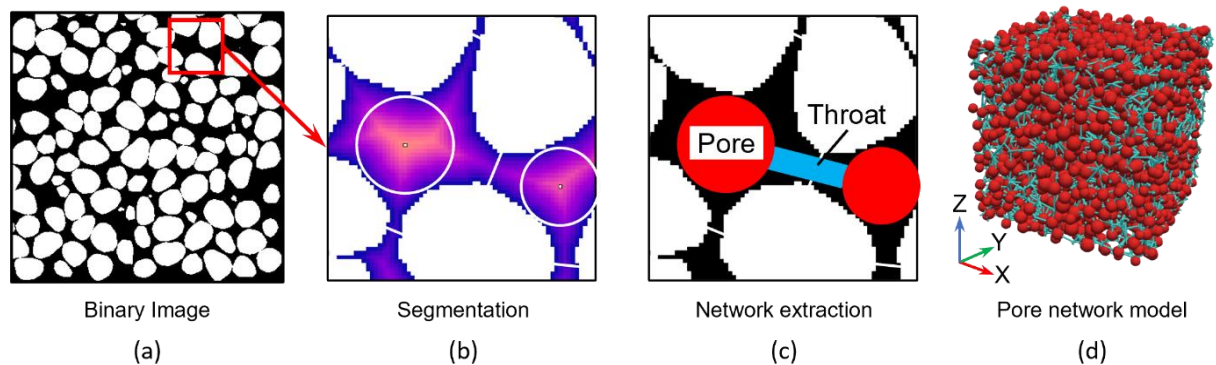

*Figure 4 Steps in Constructing Pore Network Models: (a) Binary image of the granular assembly, (b) Segmentation using the SNOW algorithm, (c) Representation of pores and throats, and (d) Three-dimensional pore network model.*

## 2.5 Complex network theory

A. Complex network theory

The internal connectivity and complex interactions of pore space present challenges in understanding fluid flow behaviours through granular assemblies, especially when particle shapes vary. Complex network theory examines the structure, dynamics, and functions of networks, providing tools to understand the structure, conduction, and transport behaviours within these networks. To analyse the pore structure using complex network theory, it is essential to represent the pore network model as a graph. In this graph, nodes represent the pores, and edges represent the throats connecting these pores (Figure 5 a). This representation can be achieved using network analysis libraries such as NetworkX in Python (Hagberg et al., 2008). By default, the weights on the edges are set to 1; however, when considering a physical network, weights are assigned to edges to represent the physical properties of the network.

In this research, the local hydraulic conductance of the pore-throat-pore connections was assigned to each edge to construct weighted complex networks. Local hydraulic conductance was calculated based on a pore-throat-pore model, which simplifies the geometry of connecting pores and throats into cylindrical representations (Figure 5 b). Notablely, two adjacent pores can be connected through multiple throats. The hydraulic conductance of each element was computed from the Hagen-Poiseuille equation as follows:

$$Q = C_{pk}\Delta p = \frac{\pi r_{pk}^4}{8\eta L_{pk}}\Delta p, k = 1,2, \tag{4}$$

$$Q = C_{ti}\Delta p = \frac{\pi r_{ti}^4}{8\eta L_{ti}}\Delta p, i = 1, \ldots, N_t, \tag{5}$$

where $Q$ is the volumetric flow rate, $k$ are IDs of the two connecting pores, $C_{pk}$ is the conductance of conduit $k$, $r_{pk}$ is the equivalent radius of pore $k$, $\eta$ the fluid dynamic viscosity, $L_{pk}$ is the length of the pore $k$ derived from the poer volume and equivalent diameter, $\Delta p$ is the pressure difference between pores driving the flow, $C_{ti}$ is the conductance of the conduit $i$, analogous to $C_{pk}$ but for a different set of conduits, $r_{ti}$ is the equivalent radius of the throat $i$ based on its cross-sectional area, $L_{ti}$ is the length of conduit $i$, $N_t$ the total number of throats connecting the two pores.

The equations for $L_{teqv}$ and $C_{teqv}$ are used to compute the equivalent length and equivalent conductance of pore throats in a network. In these equations:

$$L_{teqv} = \frac{\sum_{i=1}^{N_t} L_{ti} A_{ti}}{\sum_{i=1}^{N_t} A_{ti}} \tag{6}$$

$$C_{teqv} = \frac{\sum_{i=1}^{N_t} C_{ti} A_{ti}}{\sum_{i=1}^{N_t} A_{ti}} \tag{7}$$

where $L_{teqv}$ is the arithmetic mean of the throat lengths weighted by cross-sectional areas, $C_{teqv}$ the arithmetic means of the throat conductance weighted by their cross-sectional areas, $L_{ti}$ length of each throat, $A_{ti}$ area of each throat, $N_t$ the total number of throats, $C_{ti}$ the conductance of the i' th throat.

The overall local hydraulic conductance is calculated as:

$$C = \frac{L_{p1} + L_{teqv} + L_{p2}}{\frac{L_{p1}}{C_{p1}} + \frac{L_{teqv}}{C_{teqv}} + \frac{L_{p2}}{C_{p2}}} \tag{8}$$

where $C$ the conductance of the pore-throat-pore system, $L_{p1}$ the length of pore $p1$, $L_{p2}$ the length of pore $p2$, $L_{teqv}$ is the equivalent length of the throats.

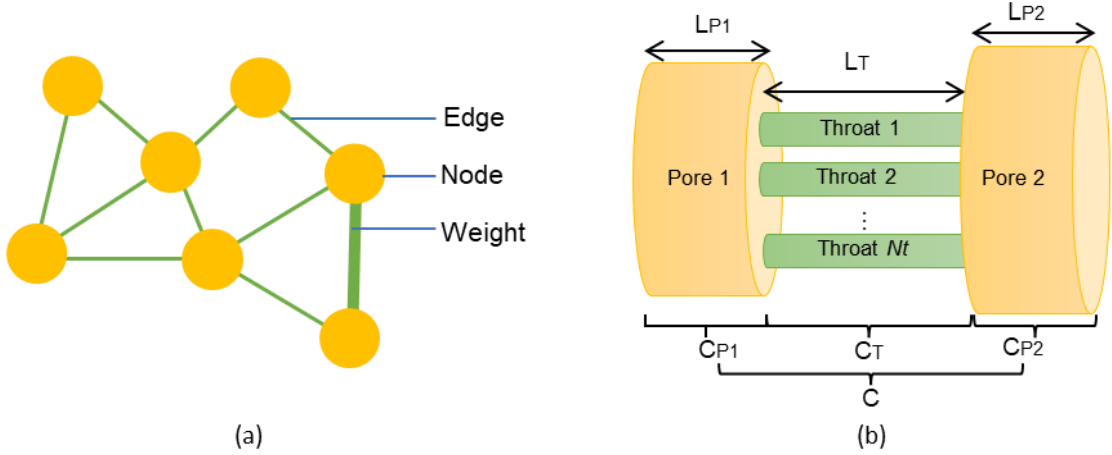

*Figure 5 (a) complex network model, (b) equivalent pore-throat-pore conduit to calculate local hydraulic conductance.*

B. Network features

Once the pore network is represented as a graph, various network properties can be analysed to gain insights into its structure and functionality. This approach allows for a comprehensive understanding of the network's connectivity and potential bottlenecks, influencing fluid flow dynamics and overall network efficiency. Key properties include:

*Degree:* The degree distribution examines the distribution of connections (throats) per pore (node). It provides critical information on the connectivity and potential bottlenecks within the network, which is essential for understanding how fluid flows through the network.

*Closeness Centrality:* Closeness centrality quantifies the relative proximity of a node to all other nodes in the network. It is computed as the reciprocal of the sum of the shortest path distances from that node to every other node. A node with a high closeness centrality is considered more centrally located, allowing it to interact with other nodes more efficiently. (Figure 6 a).

*Betweenness Centrality:* Betweenness centrality is a metric that highlights essential connectors within the network. It is defined by the proportion of shortest paths in the network that pass through a specific node. Flow through the network relies largely on the nodes with high betweenness centrality, their removal can significantly impact the network's overall connectivity (Figure 6 b).

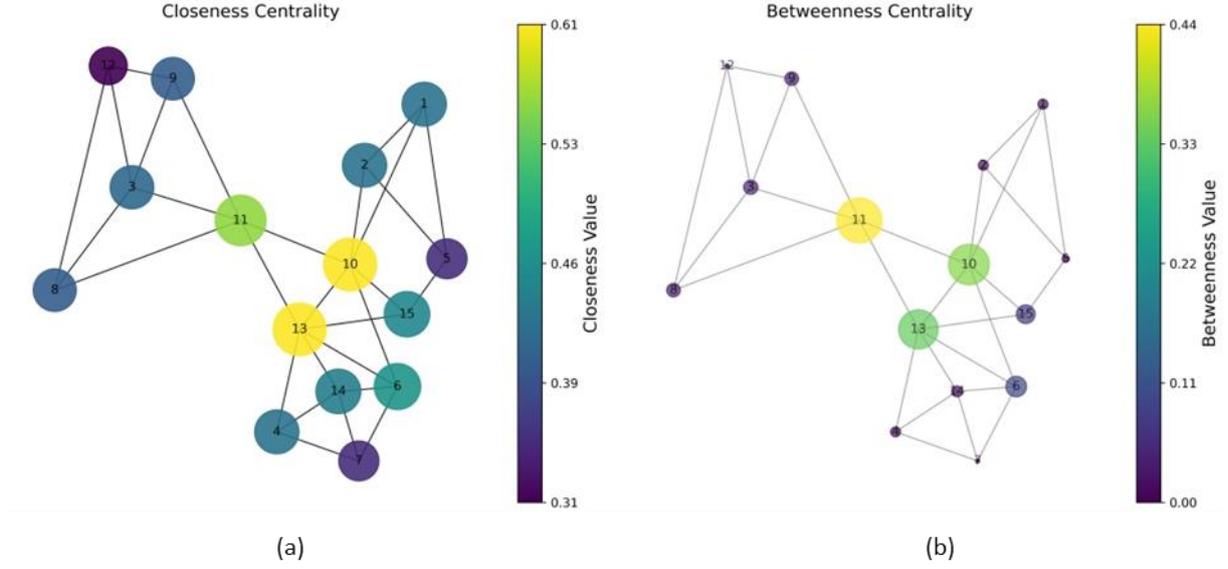

*Figure 6 Complex network features: (a) closeness centrality, (b)betweenness centrality.*

## 2.6 Hydraulic properties based on pore network model

A. Permeability

The permeability of a porous material quantifies its ability to transmit fluids under a pressure gradient. In this study, permeability is calculated using the pore network model (PNM) extracted from 3D binary images of granular assemblies. By simulating steady-state fluid flow through the pore network, we determine the volumetric flow rate and use Darcy's law to calculate the permeability of the material:

$$K = \frac{Q \cdot L}{A \cdot \Delta P} \quad (9)$$

where $K$ the permeability, $Q$ the volumetric flow rate through the sample, $L$ length of the sample in the direction of flow, $A$ cross-sectional area perpendicular to the flow direction, $\Delta P$ pressure difference applied across the sample.

The volumetric flow rate between two connected pores is calculated based on the hydraulic conductance and the pressure difference between the pores:

$$Q_{ij} = C_{ij}(P_i - P_j) \quad (10)$$

$$Q = \sum_{(i,j) \in outlet} Q_{ij} \quad (11)$$

where $Q_{ij}$ the flow rate between pores $i$ and $j$, $C_{ij}$ the hydraulic conductance the conduit between pores $i$ and $j$, $P_i$ and $P_j$ the pressures at pores $i$ and $j$.

B. Tortuosity

Tortuosity is a dimensionless parameter that characterizes the complexity of fluid flow paths in a porous medium. It provides a measure of how much longer the actual flow paths are compared to the straight-line distance through the material. A higher tortuosity indicates more convoluted paths, which can significantly impact the fluid transport properties of the material.

The tortuosity is defined as the ratio between the effective flow path length traced through the pore network which represent the actual paths taken by fluid flow, and the direct length between the inlet and outlet:

$$T = \frac{L_e}{L_d} \tag{12}$$

where $T$ the tortuosity, $L_e$ the effective flow path length through the pore network, $L_d$ the direct length between the inlet and outlet.

The effective flow path length is computed as the average length of all flow paths through the pore network. The calculation is weighted by the flow rate through each throat, meaning that throats with higher flow rates contribute more to the overall effective path length. This ensures that the flow paths are not only geometrically based but also reflect the hydraulic properties of the network.

$$L_e = \sum_{(i,j)} \frac{Q_{ij} \cdot l_{ij}}{Q} \tag{13}$$

where $L_e$ the effective flow path length, $Q_{ij}$ the flow rate between pores $i$ and $j$, $l_{ij}$ geometric length of the throat connecting pores $i$ and $j$, $Q$ the total volumetric flow rate through the sample.

## 3 Results and analysis

The pore network model derived from the assemblies contains detailed pore-scale information, such as pore and throat equivalent diameters, local hydraulic conductance, and complex network features. This model provides comprehensive insight into the internal structure of the system, as shown in Figure 7. It facilitates deeper analysis of the statistical distributions of pore and throat sizes, throat orientation, and both local and global connectivity features.

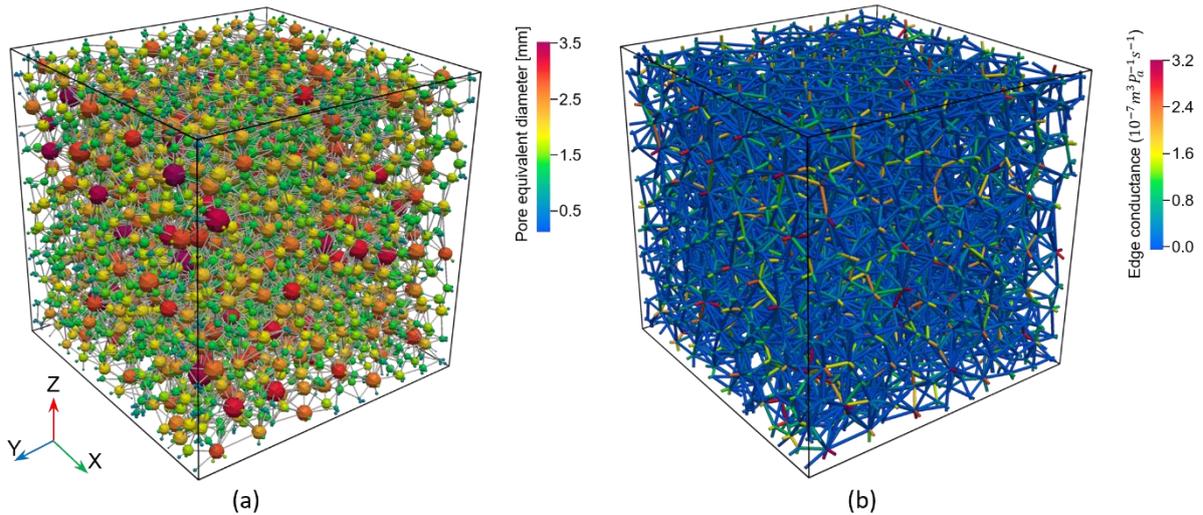

*Figure 7 Pore network model from the G1 assembly: (a) pores represented by spheres, scaled and coloured based on their equivalent diameters; (b) throats represented by cylinders, coloured by local hydraulic conductance.*

## 3.1 Pore and throat size statistical distribution

To investigate the influence of particle shapes on the pore network structures, this study employs statistical analysis of the pore and throat sizes across assemblies composed of varying particle shapes. The probability distributions of SNOW segmented pore and throat geometric parameters (*Table 3*) such as pore/throat equivalent diameter and throat equivalent length are visually represented in *Figure 8*.

The distributions reveal that particle shape induced notable variability in the distribution of these geometric parameters. Assemblies of rounded particles such as G1, generally show a tendency towards larger equivalent diameters, as indicated by the higher relative frequency of the peaks observed in the pore surface area (Figure 8a). G1 also has the largest equivalent pore diameter (Figure 8b). Specifically, the probability peaks for G0 and G1 particles at a surface area of 3-5mm$^2$ and equivalent diameter of 1.0-1.5 mm are higher than the rest particles. Other particle shapes like G4, and G5 show varying peaks but generally exhibit smaller pore and throat sizes compared to the more spherical ones. However, the distribution also shows significant overlap with other particle types, suggesting the influence of particle shape is complicated. For the throat geometric distributions (Figure 8 c & d), platy particles such as G5 tend to have a higher frequency of thinner but longer throats. This suggests that the more spherical the particle, the larger the pore and throat diameters tend to be, which could imply greater permeability and reduced resistance to fluid flow within the material. In contrast, platy particles, such as G5, display a distinctly different distribution pattern. These particles tend to feature smaller equivalent diameters, with longer equivalent throat lengths. This characteristic reflects the tendency of flatter particles to pack more tightly and align in a way that restricts pore and throat dimensions, potentially affecting the hydraulic and structural properties of the granular assembly.

These observations highlight the intricate relationship between particle morphology and the resulting geometrical properties of granular materials. Besides, this analysis provides a quantitative basis for the detailed analysis of pore network properties.

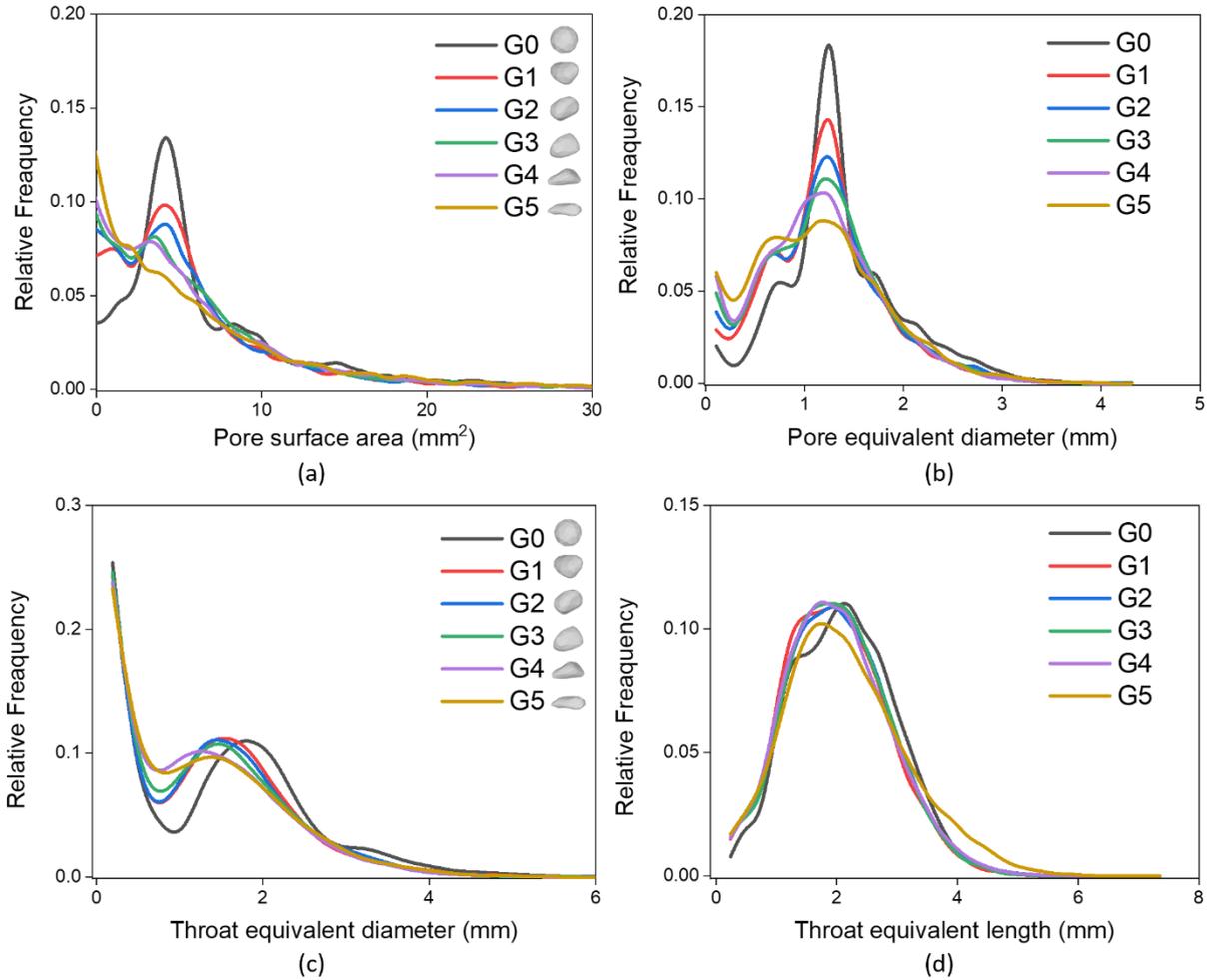

*Figure 8 Pore network size distribution:(a) pore surface area, (b) throat size distribution, (c)throat equivalent diameter, (d)throat equivalent length.*

## 3.2 Throat orientation

The orientation of the pore network is a vital factor influencing the fluid flow pathways. To further understand the impact of particle shape on pore network orientations, throat directions in the network were considered as vectors projected onto the vertical XZ and YZ planes. The probability distribution of these projected orientations was statistically calculated in a polar coordinate system (Figure 9) to uncover the patterns and anisotropy within the pore space of different particle shapes.

The relatively higher distribution of throats in the horizontal and vertical directions across different particle shapes, particularly in G0 and G1 (Figure 9 a & b), can be potentially attributed to the combined effects of boundary constraints, monodisperse nature of assemblies, and the compaction process. The compaction process within the rigid cubic box generates principal stress axes perpendicular to the walls, facilitating the preferential alignment of particles and pores along these principal directions. The monodisperse nature of the particles further promotes such configurations, where throats are more likely to align in horizontal and vertical directions. Together, these factors lead to the observed higher throat orientation distribution in the horizontal and vertical directions.

Particle shape also plays an important role in shaping the throat orientations. Rounded particles such as G0 and G1 exhibit near-uniform distributions across both horizontal and vertical directions, suggesting that rounded particles introduce minimal additional directional bias. In contrast, as particle shapes become more irregular, throat orientation distributions begin to deviate, showing a growing preference for horizontal alignment. This is particularly pronounced in platy particles like G5 (Figure 9 f), where the throat aligns predominantly along the horizontal axis.

These comparisons highlight that particle shape influences the structural anisotropy of the pore network, with platy particles introducing significant directional biases. Such variability in throat orientations could potentially induce orientation-dependent hydraulic behaviours of the granular assembly.

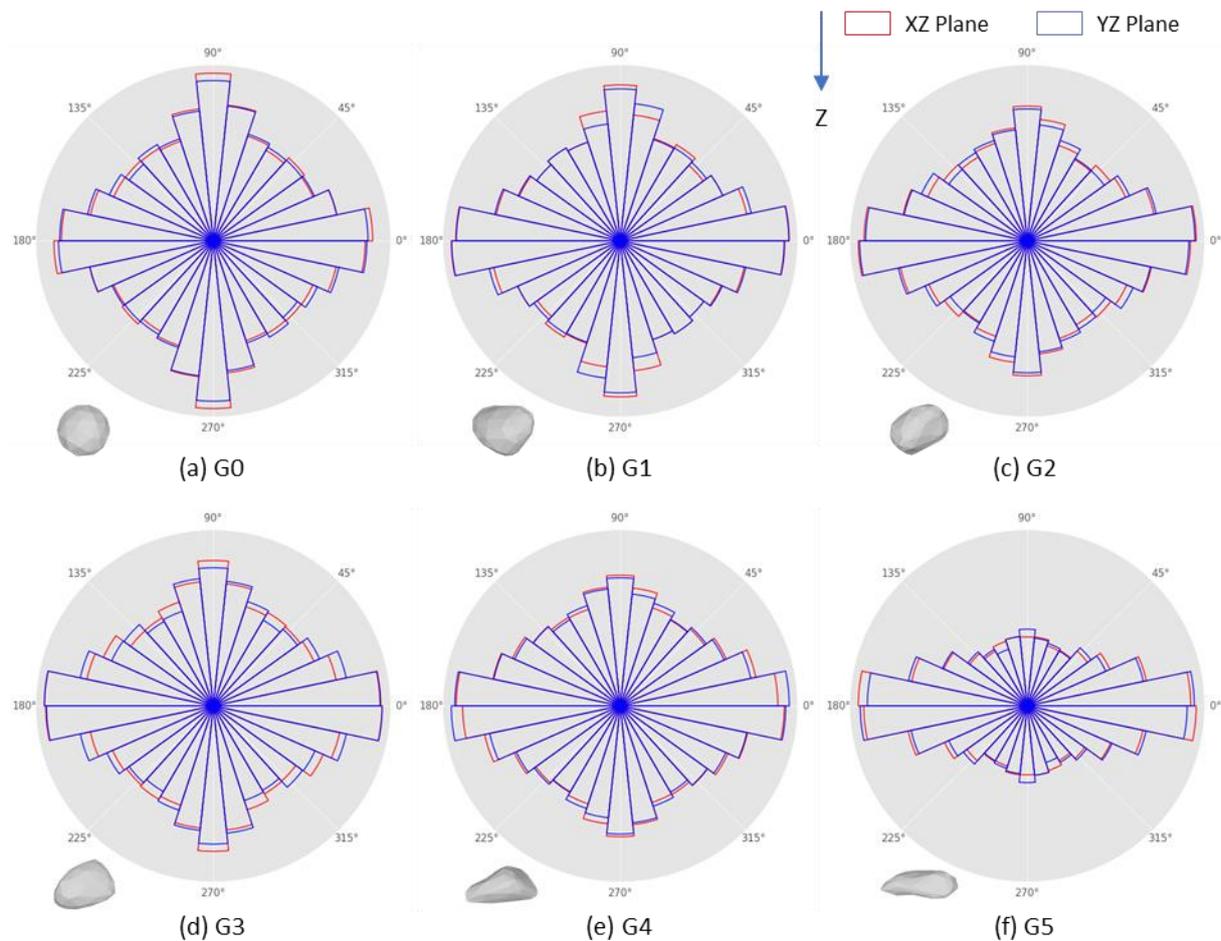

Figure 9 Orientation distribution of throats for different particle shapes in XZ and YZ planes

## 3.3 Pore network local connectivity

The connectivity between adjacent pores in the pore network model is vital for quantifying the fluid flow behaviours through porous media. The degree of pores is defined as the number of connections to adjacent pores through throats. The statistical distribution of pore degrees was hereby calculated from the irregular assemblies to reveal the distribution of localised pore connectivity Figure 10.

The statistical distribution pore degrees exhibit characteristics of power-law distributions or exponential decay distributions. A high frequency of nodes with low degrees is observed, indicating that a large portion of nodes have only a few connections. The tail of the distribution gradually decreases as the degree increases, suggesting that there are fewer nodes with higher degrees. This implies that while most pores in the network have limited connectivity, there are still significant pores that act as critical junctions with higher degrees, potentially playing a crucial role in fluid transport and network stability.

Particle shape has induced deviation in the connectivity of pore networks. This influence is evident when comparing the degree distributions of different assemblies. For instance, the peak degree in the G1 assembly is around 5, whereas the peak degree in the G5 assembly is around 8. This shift suggests that variations in particle shape led to changes in the network's connectivity. Rounded particles (G1) tend to form networks where pores have moderate connectivity, while irregular particles (G5) lead to networks with more highly connected pores. This can be attributed to the packing density and arrangement of particles, where flatter or more irregular shapes can create more complex and interlinked pore structures.

The analysis of pore degree distributions provides valuable insights into the connectivity and overall structure of pore networks in granular assemblies. The observed variations due to particle shape highlight the complex interplay between geometry and material properties.

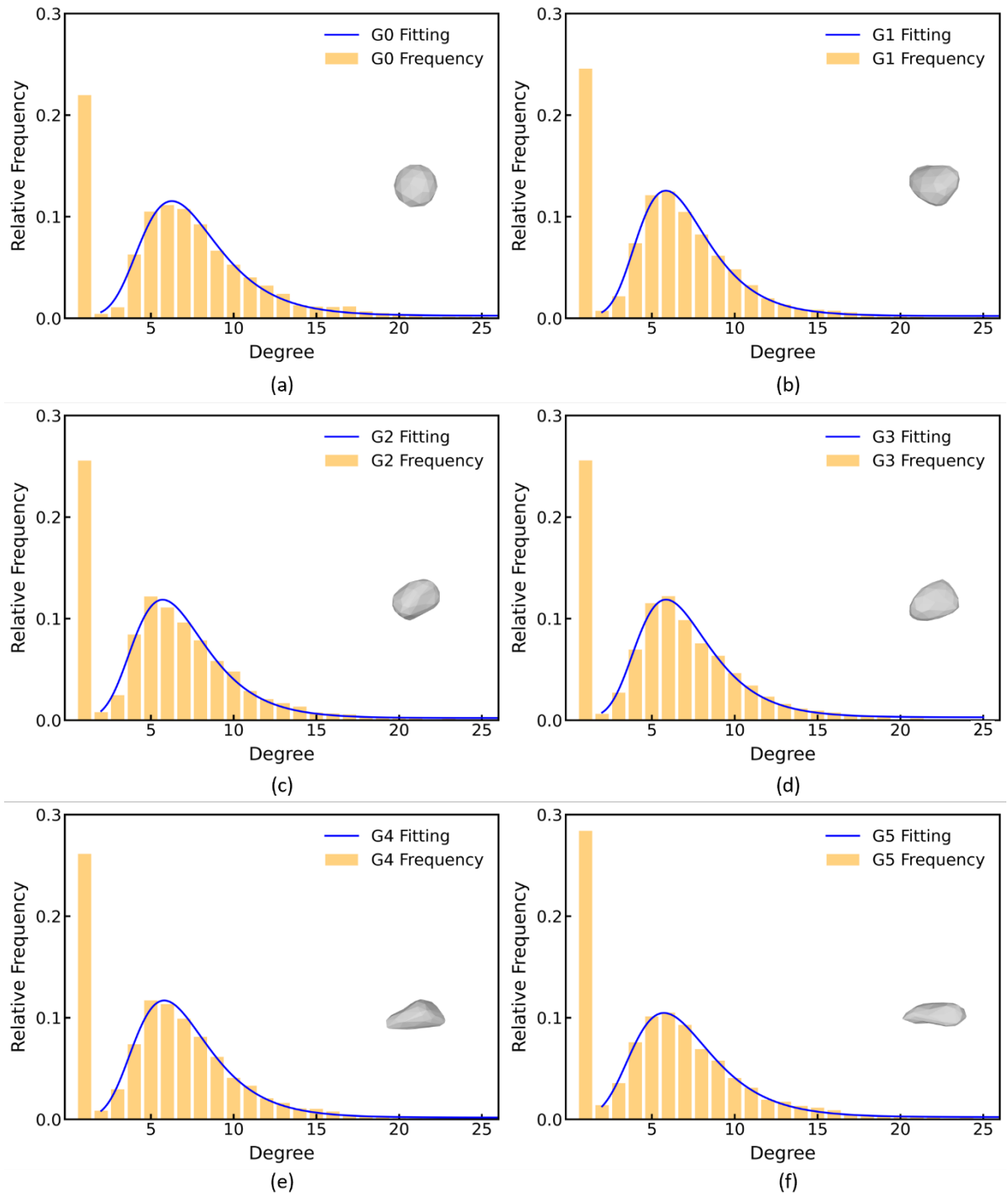

*Figure 10 Relative frequency of pore degrees.*

### 3.4 Pore network global connectivity features

To quantify the global connectivity of the constructed pore networks, network features are calculated following complex network theory. Specifically, the closeness centrality (Cc) and betweenness centrality of nodes are computed using NetworkX and visualised in ParaView to gain insights into the structural properties (Figure 11).

Closeness centrality measures how close a node is to all other nodes in the network, indicating regions where nodes can efficiently reach others. For the three particle shapes (Figure 11 a-c), regions with higher closeness centrality are concentrated in central areas, indicating efficient connectivity. The differences in these patterns across shapes suggest that particle geometry influences the ease of connectivity within the network. Betweenness centrality measures the extent to which a node lies on the shortest paths between other nodes, highlighting critical bridges within the network. The distribution of high betweenness nodes appears more scattered compared to closeness centrality (Figure 11 d-f), emphasizing specific paths that are essential for network integrity. The spatial distribution of these high betweenness nodes varies with particle shape, reflecting the impact of geometry on the network's structural and transport properties. The variations in the distribution of high centrality nodes between different particle shapes indicate that particle shape plays a significant role in determining the efficiency and critical paths within the pore network.

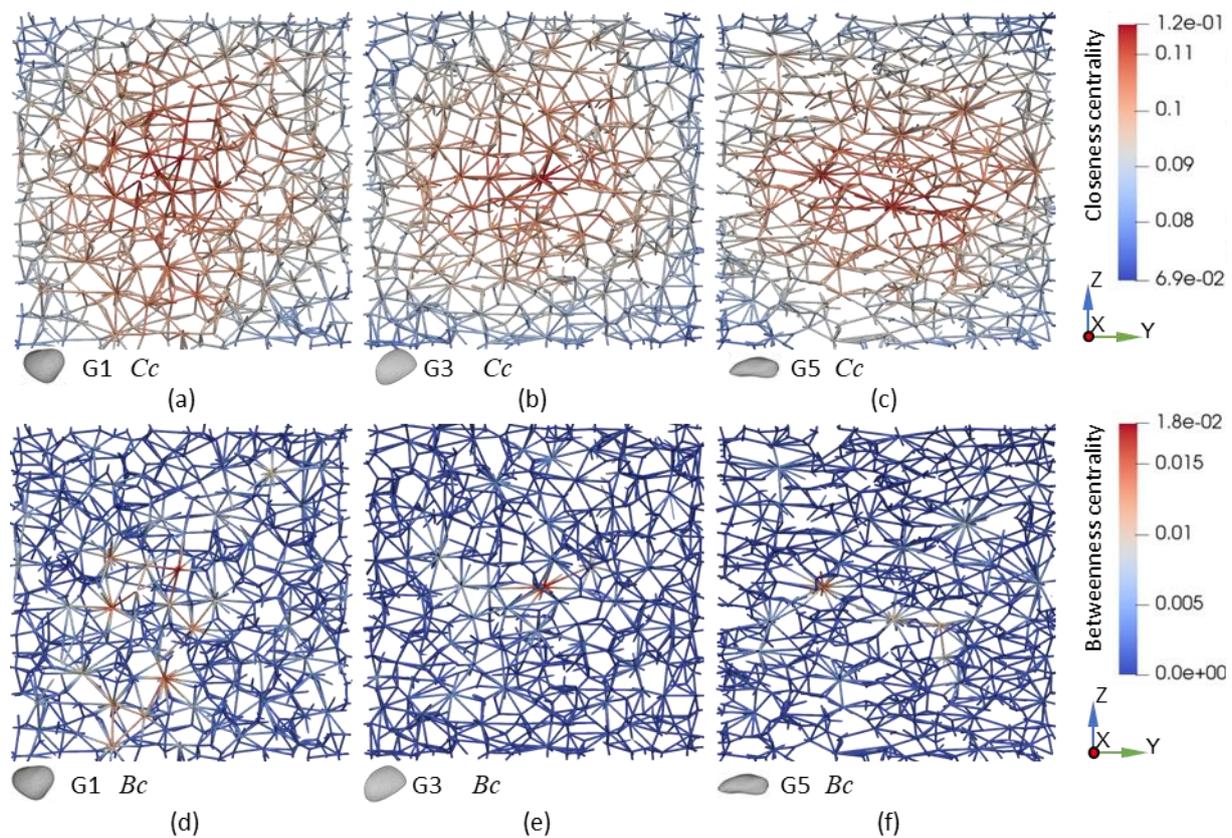

*Figure 11 Spatial distribution of pore network features: (a)-(c) closeness centrality of G1, G3 and G5 assembly, (d)-(e) betweenness centrality of G1, G3 and G5 assembly. Nodes hidden for better visualisation.*

The node closeness centrality for each pore and their probability distribution are statistically calculated, as depicted in Figure 12. The distribution of betweenness centrality (Figure 12a) exhibits minor variation induced by particle shape, indicating that while particle morphology slightly affects the local betweenness centrality distribution, it does not significantly alter the overall betweenness connectivity distributions. The closeness centrality distributions for different particle shapes are relatively similar (Figure 12b), suggesting that the geometric arrangement of particles has a limited impact on the network's overall closeness centrality. This implies that the inherent connectivity of the pore network remains relatively stable across different particle shapes when considering the unweighted pore network model alone.

The closeness centrality has been proved to have a strong correlation with the hydraulic properties of sphere assemblies (van der Linden et al., 2018), and thus was employed to characterise the pore network structures. To link the pore networks with the macroscopic hydraulic properties, weighted closeness centrality was calculated from the local hydraulic conductance weighted pore network model (Figure 12c). It is observed that the particle shape has a significant impact on the weighted closeness centrality distribution. Networks formed by rounded particles, such as G1, exhibit higher weighted closeness centrality compared to those formed by platy particles, such as G4 and G5. This indicates that networks with rounded particles facilitate more efficient pathways for fluid flow. The higher weighted closeness centrality observed in these networks suggests shorter and more direct routes between pores, enhancing hydraulic conductivity. Conversely, the lower weighted closeness centrality in networks formed by platy particles implies more tortuous pathways, potentially leading to decreased hydraulic conductivity.

This variation underscores the critical role of particle morphology in determining the efficiency of fluid transport through the pore network. The more complex and indirect routes associated with platy particles may hinder fluid movement, resulting in less efficient hydraulic performance.

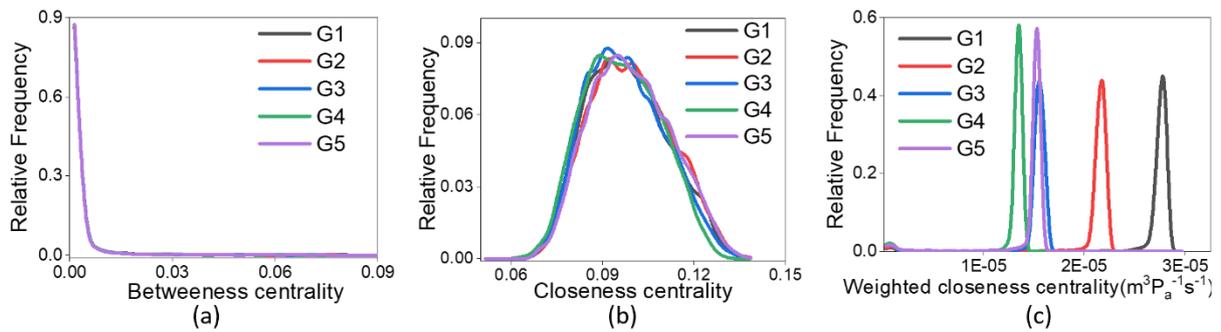

*Figure 12 Probability distribution of pore network features: (a) closeness centrality (b)hydraulic conductance weighted closeness centrality*

## 3.5 Hydraulic properties

The influence of particle shape on the hydraulic properties is further investigated focusing on permeability. In this study, a pressure difference of 100 Pa was applied along the X and Z directions separately for each granular assembly to simulate fluid flow through the pore networks using Stokes flow equations integrating hydraulic conductance from the pore-throat-pore conduit (Figure 13a). The permeability of each assembly in each direction was then calculated using Darcy's law. To facilitate the comparison of the results across different particle sizes, the permeability values were normalized by dividing the square of the particle equivalent diameter $d_e$. The normalized permeability of the spherical assembly was compared with established experimental (Zick & Homsy, 1982) and numerical (Zhao & O'Sullivan, 2022) benchmark results on spherical packings (Figure 13b). The agreement between the calculated permeability and the benchmarks demonstrates the reliability of the pore network model in capturing the hydraulic behaviour of the granular assemblies.

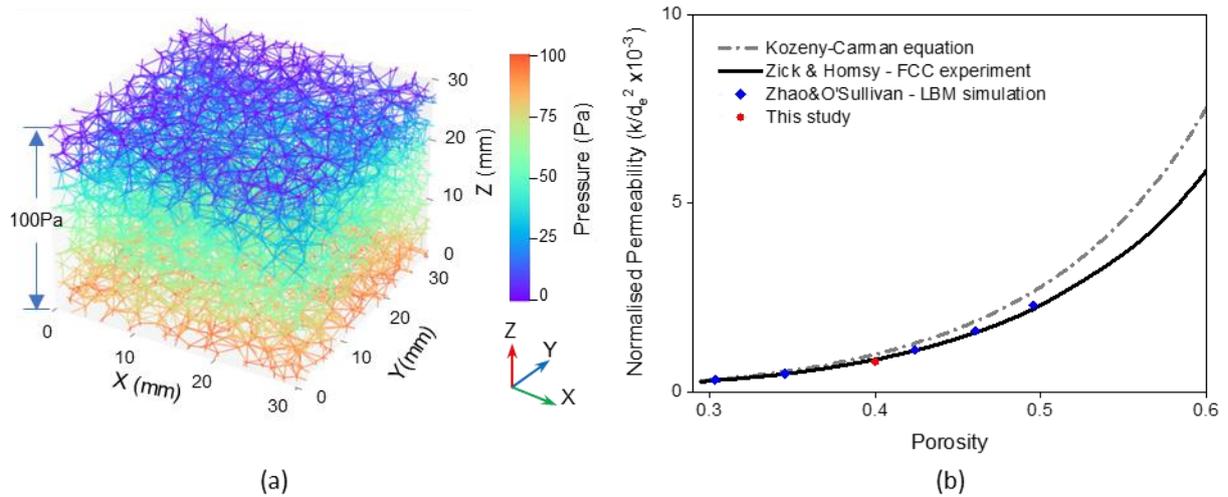

*Figure 13 Permeability analysis from the pore network model: (a) pressure distribution on the G0 pore network model with pores hidden for better visualization, (b) validation of permeability for the spherical G0 assembly.*

The permeability of the particle shapes with different sphericity at the X and Z directions are compared in (Figure 14a) The permeability in different directions demonstrates the impact of particle shape on hydraulic anisotropy. For rounded particles such as G1 and G2, the differences between permeability in the X and Z directions (Kx and Kz) are minimal, indicating more uniform hydraulic behaviours in horizontal and vertical directions. This is consistent with the relatively uniform distribution of throat orientation in (Figure 9a and b).

As sphericity decreases, the difference between Kx and Kz gradually increases. For particles with lower sphericity (e.g., platy G5 particle), the permeability in the Z direction (Kz) is significantly lower than in the X direction (Kx), indicating strong anisotropy in the flow. This behaviour can be attributed to the preferential horizontal alignment of platy particles during compaction, which enhances horizontal pore network connectivity as evidenced by the throat orientation distribution (Figure 9f). These findings highlight that particle shape plays a critical role in the anisotropic fluid flow behaviours in granular assemblies.

Tortuosity quantifies the extent of distortion in flow paths along the macroscopic flow direction and was calculated from the effective flow path length divided by the straight-line distance along the specific direction (Figure 14 (b)). The findings from Figure 14 (b) show that tortuosity in the horizontal direction (Tx) and vertical direction (Tz) is also influenced by particle shape. Platy particles with lower sphericity show higher tortuosity, especially in the vertical direction. This increase in tortuosity correlates with the more complex flow paths created by the irregular shapes of these particles.

The significant anisotropy observed in both permeability and tortuosity underscores the pivotal role of particle morphology in determining the hydraulic behaviour of these materials. Platy particles, characterised by lower sphericity, induce greater anisotropy, resulting in higher directional variability in both permeability and tortuosity. This indicates that the shape and arrangement of particles can significantly alter the efficiency and directionality of fluid transport within the material. Understanding these effects is crucial for predicting and optimizing the hydraulic performance of granular materials in various engineering applications.

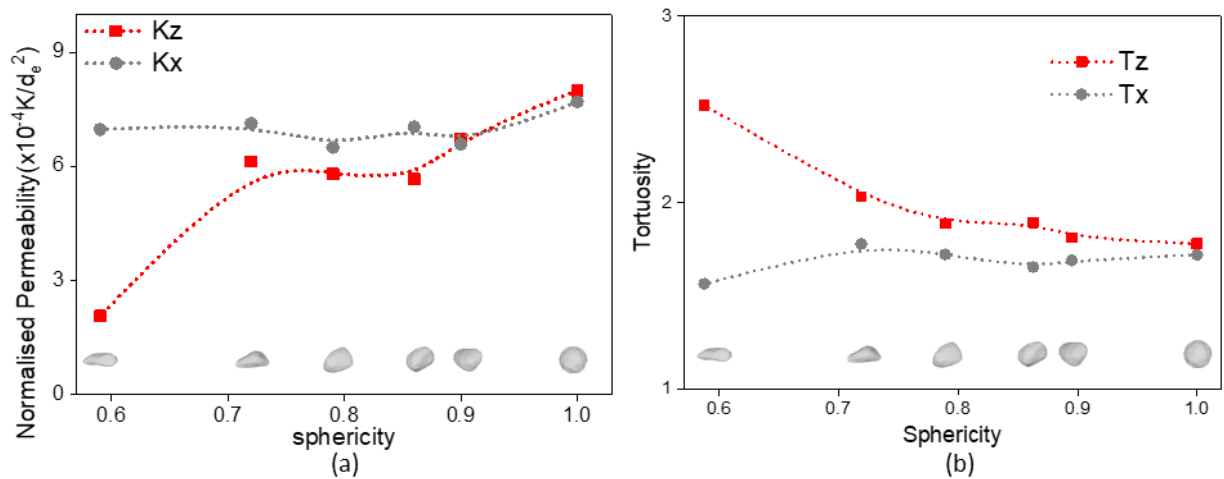

*Figure 14 Influence of particle shape on hydraulic properties (a) normalised permeability (b) tortuosity*

## 4 Conclusions

This study utilizes a combination of DEM simulation, image processing, and complex network analysis to explore the impact of particle shapes on the pore network and hydraulic properties of monodisperse granular assemblies. The key findings and insights from this comprehensive analysis are as follows:

The pore structures of granular assemblies are intrinsically influenced by the shapes of the constituent particles. Assemblies composed of more rounded particles generally exhibit larger pore and throat sizes, which facilitate better connectivity within the pore network. This enhanced connectivity can lead to increased permeability and fluid transport efficiency. This insight underscores the fundamental role of particle morphology in dictating the spatial arrangement and connectivity of pores.

Structural and hydraulic anisotropy induced by particle shape is observed, particularly pronounced in platy particles such as G5. The anisotropic behaviour observed in assemblies with platy particles highlights how particle shape can influence not only the microstructure but also the macroscopic hydraulic properties. The relationship between particle shape, permeability, and tortuosity underscores the intricate interplay between microstructural characteristics and fluid flow behaviour. Recognizing and leveraging this relationship is essential for advancing both the theoretical understanding and practical applications of granular materials in engineering and environmental contexts. For instance, this knowledge can be vital in designing more effective drainage systems, improving groundwater management, and ensuring the stability of soil structures under various environmental conditions.

Weighted edge closeness centrality effectively reflects the influence of particle shape on pore structures and hydraulic properties. Complex network features provide an in-depth understanding of how the proximity and connectivity of pores within the network influence fluid transport. By linking closeness centrality to hydraulic conductance, the study bridges the gap between network topology and practical hydraulic performance, offering a valuable tool for predicting and optimizing the flow and transport behaviours in granular material.

While this study focuses on the influence of particle shape in monodisperse assemblies, it is crucial to address the lack of consideration for the heterogeneity and polydispersity of real-world granular materials. Future research could expand the investigation to include a broader range of particle shapes and sizes, as well as more detailed network features, to gain deeper insights into pore network dynamics. Additionally, incorporating dynamic simulations and more complex boundary conditions will help address the limitations associated with the assumption of static pore structures, which may not fully capture the dynamic nature of real-world conditions.

## 5 Acknowledgements

The authors gratefully acknowledge the financial support from the Australian Research Council under project ARC DP210100433 and the Chinese Scholarship Council. The authors sincerely appreciate the High-Performance Computing (HPC) service provided by The University of Melbourne Research Computing Portal (RCP).

# 6 References

Abràmoff, M. D., Magalhães, P. J., & Ram, S. (2004). Image Processing with ImageJ. *Biophotonics International*, *11*(7), 36–42.

Ahrens, J., Geveci, B., & Law, C. (2005). ParaView: An End-User Tool for Large-Data Visualization. In *Visualization Handbook* (pp. 717–731). Elsevier. https://doi.org/10.1016/B978-012387582-2/50038-1

Alonso, M., Vaunat, J., Vu, M. N., Talandier, J., Olivella, S., & Gens, A. (2024). Three-Dimensional Modelling of a Large-Diameter Sealing Concept in a Deep Geological Radioactive Waste Disposal. *Rock Mechanics and Rock Engineering*. https://doi.org/10.1007/s00603-024-03813-w

Bidarmaghz, A., Makasis, N., Fei, W., & Narsilio, G. A. (2021). An efficient and sustainable approach for cooling underground substations. *Tunnelling and Underground Space Technology*, *113*, 103986. https://doi.org/10.1016/j.tust.2021.103986

Blain, J. M. (2019). *The Complete Guide to Blender Graphics*. A K Peters/CRC Press. https://doi.org/10.1201/9780429196522

Blunt, M. J., Bijeljic, B., Dong, H., Gharbi, O., Iglauer, S., Mostaghimi, P., Paluszny, A., & Pentland, C. (2013). Pore-scale imaging and modelling. *Advances in Water Resources*, *51*, 197–216. https://doi.org/10.1016/j.advwatres.2012.03.003

Cignoni, P., Callieri, M., Corsini, M., Dellepiane, M., Ganovelli, F., & Ranzuglia, G. (2008). *MeshLab: an Open-Source Mesh Processing Tool*.

Ding, W. T., & Xu, W. J. (2018). Study on the multiphase fluid-solid interaction in granular materials based on an LBM-DEM coupled method. *Powder Technology*, *335*, 301–314. https://doi.org/10.1016/j.powtec.2018.05.006

Fei, W. Bin, Li, Q., Wei, X. C., Song, R. R., Jing, M., & Li, X. C. (2015). Interaction analysis for CO2 geological storage and underground coal mining in Ordos Basin, China. *Engineering Geology*, *196*, 194–209. https://doi.org/10.1016/j.enggeo.2015.07.017

Fei, W., & Narsilio, G. A. (2020). *Impact of Three-Dimensional Sphericity and Roundness on Coordination Number*. https://doi.org/10.1061/(ASCE)

Fei, W., Narsilio, G. A., & Disfani, M. M. (2019). Impact of three-dimensional sphericity and roundness on heat transfer in granular materials. *Powder Technology*, *355*, 770–781. https://doi.org/10.1016/j.powtec.2019.07.094

Fei, W., Narsilio, G. A., van der Linden, J. H., Tordesillas, A., Disfani, M. M., & Santamarina, J. C. (2021). Impact of particle shape on networks in sands. *Computers and Geotechnics*, *137*. https://doi.org/10.1016/j.compgeo.2021.104258

Galindo-Torres, S. A. (2013). A coupled Discrete Element Lattice Boltzmann Method for the simulation of fluid–solid interaction with particles of general shapes. *Computer Methods in Applied Mechanics and Engineering*, *265*, 107–119. https://doi.org/10.1016/j.cma.2013.06.004

Garland, M., & Heckbert, P. S. (1997). *Surface Simplification Using Quadric Error Metrics*. http://www.cs.cmu.edu/

Gerke, K. M., Sizonenko, T. O., Karsanina, M. V., Lavrukhin, E. V., Abashkin, V. V., & Korost, D. V. (2020). Improving watershed-based pore-network extraction method using maximum inscribed ball pore-body positioning. *Advances in Water Resources*, *140*. https://doi.org/10.1016/j.advwatres.2020.103576

Gostick, J., Aghighi, M., Hinebaugh, J., Tranter, T., Hoeh, M. A., Forschungszentrum, |, Day, J. H., Spellacy, B., Sharqawy, M. H., Burns, A., Lehnert, W., Jülich, F., Aachen, R., & Putz, A. (2016). *OpenPNM: A Pore Network Modeling Package*. www.scipy.org

Gostick, J., Khan, Z., Tranter, T., Kok, M., Agnaou, M., Sadeghi, M., & Jervis, R. (2019). PoreSpy: A Python Toolkit for Quantitative Analysis of Porous Media Images. *Journal of Open Source Software*, *4*(37), 1296. https://doi.org/10.21105/joss.01296

Gostick, J. T. (2017). Versatile and efficient pore network extraction method using marker-based watershed segmentation. *Physical Review E*, *96*(2). https://doi.org/10.1103/PhysRevE.96.023307

Icardi, M., Boccardo, G., Marchisio, D. L., Tosco, T., & Sethi, R. (2014). Pore-scale simulation of fluid flow and solute dispersion in three-dimensional porous media. *Physical Review E - Statistical, Nonlinear, and Soft Matter Physics*, *90*(1). https://doi.org/10.1103/PhysRevE.90.013032

Kantzas, A., Bryan, J. L., Mai, A., & Hum, F. M. (2005). Applications of low field NMR techniques in the characterization of oil sand mining, extraction and upgrading processes. *Canadian Journal of Chemical Engineering*, *83*(1), 145–150. https://doi.org/10.1002/cjce.5450830124

Li, Z., Galindo-Torres, S., Yan, G., Scheuermann, A., & Li, L. (2019). Pore-Scale Simulations of Simultaneous Steady-State Two-Phase Flow Dynamics Using a Lattice Boltzmann Model: Interfacial Area, Capillary Pressure and Relative Permeability. *Transport in Porous Media*, *129*(1), 295–320. https://doi.org/10.1007/s11242-019-01288-w

Liang, Y., Hu, P., Wang, S., Song, S., & Jiang, S. (2019). Medial axis extraction algorithm specializing in porous media. *Powder Technology*, *343*, 512–520. https://doi.org/10.1016/j.powtec.2018.11.061

Lindquist, W. B., Lee, S. M., Coker, D. A., Jones, K. W., & Spanne, P. (1996). Medial axis analysis of void structure in three-dimensional tomographic images of porous media. *Journal of Geophysical Research: Solid Earth*, *101*(4), 8297–8310. https://doi.org/10.1029/95jb03039

Lv, J., Cheng, Z., Xue, K., Liu, Y., & Mu, H. (2020). Pore-scale morphology and wettability characteristics of xenon hydrate in sand matrix - Laboratory visualization with micro-CT. *Marine and Petroleum Geology*, *120*. https://doi.org/10.1016/j.marpetgeo.2020.104525

Minagawa, H., Nishikawa, Y., Ikeda, I., Miyazaki, K., Takahara, N., Sakamoto, Y., Komai, T., & Nairta, H. (2008). Characterization of sand sediment by pore size distribution and permeability using proton nuclear magnetic resonance measurement. *Journal of Geophysical Research: Solid Earth*, *113*(7). https://doi.org/10.1029/2007JB005403

Mohamad, A. A. (2019). *Lattice Boltzmann Method*. Springer London. https://doi.org/10.1007/978-1-4471-7423-3

Mukunoki, T., Miyata, Y., Mikami, K., & Shiota, E. (2016). X-ray CT analysis of pore structure in sand. *Solid Earth*, *7*(3), 929–942. https://doi.org/10.5194/se-7-929-2016

Prakongkep, N., Suddhiprakarn, A., Kheoruenromne, I., & Gilkes, R. J. (2010). SEM image analysis for characterization of sand grains in Thai paddy soils. *Geoderma*, *156*(1–2), 20–31. https://doi.org/10.1016/j.geoderma.2010.01.003

Rabbani, A., Ayatollahi, S., Kharrat, R., & Dashti, N. (2016). Estimation of 3-D pore network coordination number of rocks from watershed segmentation of a single 2-D image. *Advances in Water Resources*, *94*, 264–277. https://doi.org/10.1016/j.advwatres.2016.05.020

Randolph, M. F., Gaudin, C., Gourvenec, S. M., White, D. J., Boylan, N., & Cassidy, M. J. (2011). Recent advances in offshore geotechnics for deep water oil and gas developments. *Ocean Engineering*, *38*(7), 818–834. https://doi.org/10.1016/j.oceaneng.2010.10.021

Roy, N., Frost, J. D., & Roozbahani, M. M. (2023). Quantifying three-dimensional bodies and throats of particulate system pore space. *Powder Technology*, *415*. https://doi.org/10.1016/j.powtec.2022.118160

Schroeder, W., Kenneth M. Martin, & William E. Lorensen. (1998). *The Visualization Toolkit: An Object-Oriented Approach to 3D Graphics*.

Sufian, A., Russell, A. R., & Whittle, A. J. (2019). Evolving pore orientation, shape and size in sheared granular assemblies. *Granular Matter*, *21*(1). https://doi.org/10.1007/s10035-018-0856-4

Sun, W. C., Kuhn, M. R., & Rudnicki, J. W. (2013). A multiscale DEM-LBM analysis on permeability evolutions inside a dilatant shear band. *Acta Geotechnica*, *8*(5), 465–480. https://doi.org/10.1007/s11440-013-0210-2

Tartakovsky, A. M., Tartakovsky, D. M., Scheibe, T. D., & Meakin, P. (2008). Hybrid simulations of reaction-diffusion systems in porous media. *SIAM Journal on Scientific Computing*, *30*(6), 2799–2816. https://doi.org/10.1137/070691097

Taylor, H. F., O'Sullivan, C., & Sim, W. W. (2015). A new method to identify void constrictions in micro-CT images of sand. *Computers and Geotechnics*, *69*, 279–290. https://doi.org/10.1016/j.compgeo.2015.05.012

Van der Linden, J. H. (2019). *Pore-scale characterisation of fluid flow and heat transfer in granular geomaterials* [PhD thesis]. University of Melbourne.

Van Der Linden, J. H., Narsilio, G. A., & Tordesillas, A. (2016). Machine learning framework for analysis of transport through complex networks in porous, granular media: A focus on permeability. *Physical Review E*, *94*(2), 1–16. https://doi.org/10.1103/PhysRevE.94.022904

Van der Linden, J. H., Sufian, A., Narsilio, G. A., Russell, A. R., & Tordesillas, A. (2018). A computational geometry approach to pore network construction for granular packings. *Computers and Geosciences*, *112*, 133–143. https://doi.org/10.1016/j.cageo.2017.12.004

Zhao, B., & O'Sullivan, C. (2022). Fluid particle interaction in packings of monodisperse angular particles. *Powder Technology*, *395*, 133–148. https://doi.org/10.1016/j.powtec.2021.09.022

Zheng, W., Hu, X., Tannant, D. D., & Zhou, B. (2021). Quantifying the influence of grain morphology on sand hydraulic conductivity: A detailed pore-scale study. *Computers and Geotechnics*, *135*. https://doi.org/10.1016/j.compgeo.2021.104147

Zhu, Y. I., Fox, P. J., & Morris, J. P. (1999). A pore-scale numerical model for flow through porous media. *International Journal for Numerical and Analytical Methods in*

*Geomechanics*, *23*(9), 881–904. https://doi.org/10.1002/(SICI)1096-9853(19990810)23:9<881::AID-NAG996>3.0.CO;2-K

Zick, A. A., & Homsy, G. M. (1982). Stokes flow through periodic arrays of spheres. *Journal of Fluid Mechanics*, *115*, 13–26. https://doi.org/10.1017/S0022112082000627

**List of Figures**

Figure 1 General-shape granular assemblies (a) DEM particle generating; (b) assembly compaction
Figure 2 Pore geometry of general-shape granular assemblies
Figure 3 Pore image extraction (a) pore geometry; (b) greyscale images of 2D pore slices; (c) 3D binary pore image stacks
Figure 4 Steps in Constructing Pore Network Models: (a) Binary image of the granular assembly, (b) Segmentation using the SNOW algorithm, (c) Representation of pores and throats, and (d) Three-dimensional pore network model.
Figure 5 (a) complex network model, (b) equivalent pore-throat-pore conduit to calculate local hydraulic conductance.
Figure 6 Complex network features: (a) Betweenness centrality, (b) Closeness centrality.
Figure 7 Pore network model from the G1 assembly: (a) pores represented by spheres, scaled and coloured based on their equivalent diameters; (b) throats represented by cylinders, coloured by local hydraulic conductance.
Figure 8 Pore network size distribution:(a) pore surface area, (b) throat size distribution, (c)throat equivalent diameter, (d)throat equivalent length.
Figure 9 Orientation distribution of throats for different particle shapes in XZ and YZ planes
Figure 10 Relative frequency of pore degrees.
Figure 11 Spatial distribution of pore network features: (a)-(c) closeness centrality of G1, G3 and G5 assembly, (d)-(e) betweenness centrality of G1, G3 and G5 assembly. Nodes hidden for better visualisation.
Figure 12 Probability distribution of pore network features: (a) closeness centrality (b)hydraulic conductance weighted closeness centrality
Figure 13 Permeability analysis from the pore network model: (a) pressure distribution on the G0 pore network model with pores hidden for better visualization, (b) validation of permeability for the spherical G0 assembly.
Figure 14 Influence of particle shape on hydraulic properties (a) normalised permeability (b) tortuosity

**List of Tables**